\documentstyle[epsf,twocolumn,pre,aps,amssymb]{revtex}
\begin{document} 
\wideabs{    
\draft
\title{ Orientational orders of small anisotropic molecules
confined in slit pores}
\author{ Xin Zhou$^1$, Hu Chen$^2$ and Mitsumasa Iwamoto$^1$}
\address{$\ ^{1}$Department of Physical Electronics, Tokyo Institute of 
Technology, O-okayama 2-12-1, Meguro-ku, Tokyo 152-8552, Japan \\
$\ ^{2}$Department of Civil Engineering, National Univeristy of Singapore}
\date{\today}

\maketitle

\begin{abstract}
We have studied phase behavior of hard gaussian overlap molecules with small 
anisotropic parameter confined in two plane parallel structureless hard walls.
Our investigation based on standard constant-NPT Monte Carlo molecular 
simulation led us to some interesting findings. For small anisotropic 
molecules the nematic phase is instable in bulk, while, if the distance 
between the walls is small enough, an orientation-ordered phase can form. 
This result indicates that the required molecular elongation forming 
liquid-crystal phases is smaller in confinement than that in bulk. 
Considering the value of the elongation of molecules, the computed
result inplies that small molecule liquid crystals may exist in confinement.
\end{abstract}
}

\section{Introduction}
There is a great number of molecular simulations in the literature on the 
properties of liquid crystals. Among these simulations, liquid crystal 
molecules are described using anisotropic models, such as the hard gaussian 
overlap (HGO) models~\cite{Miguel2003,Miguel2001,Padilla1997}, the similar 
hard spherocylinder~\cite{Dijkstra2001,McGrother1996,Veerman1990} and 
the hard spheroidal models~\cite{Venkat1987}, and the wider used Gay-Berne 
(GB) models~\cite{Miguel2002,Miguel2002b,Gruhn1998,Gay1981}. Although the 
detail results depend on the models used, some general conculsions have been 
found: for example it is generally confirmed that liquid-crystal phases can 
form only when the anisotropic parameters of these models are greater than 
some critical values. Otherwise (i.e. for molecules with small anisotropic 
parameters), as one increases the pressure or decreases the temperature, these 
systems will freeze or crystallize before forming liquid-crystal phases. 
Many recent works focus on the macroscopic properties and phase transitions 
of models with large anisotropy parameters. For example, based on molecular 
simulation and statistical theories, global phase diagrams of hard potential 
models~\cite{McGrother1996,Veerman1990} and GB models~\cite{Miguel2002} in 
bulk have been developed. 

On the other hand, confinement in nanometer scale is found to induce phase 
transitions not observed in bulk systems and shifts in phase 
transitions~\cite{Gelb1999,Lum1999,Koga2001}. 
Recently, Gelb {\it et al.}~\cite{Gelb1999} systematically reviewed 
research trend on confined simple fluids based on statistical physics, 
molecular simulations, and carried out some experimental investigations. 
The properties of 
confined liquid crystals also attracted wide interests on both sciences and 
technologies, such as the surface anchoring effects of liquid 
crystals~\cite{Lange2002}. 
More recently, Gruhn and Schoen~\cite{Gruhn1998,Gruhn1997} studied the 
microscopic structure of confined liquid crystals based on GB models. 
So far, these researches on the bulk and confined properties of liquid 
crystals only focus on systems composed of large anisotropic molecules 
which model the 
usual liquid-crystal molecules. To the best of our knowledge, properties of 
small anisotropic molecules have shown no interests, probably becuase they 
are believed not to form orientation-ordered phases in bulk~\cite{Rigby1989}.

But there is a need to address the following question: Are 
liquid crystal phases in confinement stable even in the 
small-anisotropic-molecule 
systems? According to the knowledge on confined simple fluids, although 
these systems cannot form orientation-ordered phases in bulk, confinement 
induce new phases when these molecules are confined in nanometer-sized pores.
The answer to the above question is important not only for the sake of 
properties 
of confined fluids, but implies the possibility of existing small 
molecule liquid crystals.

In this paper, we study phase behavior of molecules confined between two plane 
parallel structureless hard walls (i. e. slit pore), we investigate 
orientation-ordered phases and their transitions on a few kinds of molecules 
with different anisotropic parameters (molecular elongation), and show that 
small anisotropic molecules can form stable orientation-ordered phase 
(nemetic phase) in confinement, though in bulk the phase does not exist.
Our results indicate that the required molecular elongation which provoke 
stable liquid-crystal phase in HGO systems decreases with confining 
conditions. Therefore we suspect that small molecules might form liquid 
crystals in confinement. Some related discussions are presented in Sec. IV. 
Simulation results are shown in Sec. III. We begin, however, in Sec. II with 
a description of the HGO model and the presentation of our simulations.

\section{Potential Model and Computational Details}
Due to a general belief that repulsive interactions are key roles to 
determine liquid cyrstal structures, some hard nonspherical 
models~\cite{Miguel2003,Miguel2001,Padilla1997,Dijkstra2001,McGrother1996,Veerman1990,Venkat1987} are used to study properties 
of liquid crystals. Among them, one of the most simplest models is the HGO 
model~\cite{Miguel2003,Miguel2001,Padilla1997,Rigby1989}. In the model, 
interaction between a pair of molecules $i$ and $j$ is given by
\begin{eqnarray}
U({\mathbf r}_{ij}, {\hat u}_{i}, {\hat u}_{j}) = 
\left\{ 
\begin{array}{@{\,}ll}
\infty, & if \ \mbox{$ r_{ij} \le \sigma({\hat r}_{ij}, {\hat u}_{i}, 
{\hat u}_{j}) $} \\
0, & if \ \mbox{$ r_{ij} > \sigma({\hat r}_{ij}, {\hat u}_{i}, 
{\hat u}_{j}) $},
\end{array}
\right.
\end{eqnarray}
where, ${\hat r}_{ij} = {\mathbf r}_{ij}/r_{ij}$ is a unit vector along the 
line joining the centers-of-mass of the molecules, ${\hat u}_{i}$ and 
${\hat u}_{j}$ are unit vectors along the principal axes of the molecules, 
$\sigma$ is dependent on the molecular orientation and explicitly given by 
\begin{eqnarray}
\sigma({\hat r}_{ij}, {\hat u}_{i}, {\hat u}_{j}) &=& \sigma_{0} \{ 1 - 
\frac{\cal X}{2} [ \frac{
({\hat r}_{ij} \cdot {\hat u}_{i} + {\hat r}_{ij} \cdot {\hat u}_{j})^{2}}
{1 + {\cal X} ({\hat u}_{i} \cdot {\hat u}_{j})} \nonumber \\
&+& \frac{
({\hat r}_{ij} \cdot {\hat u}_{i} - {\hat r}_{ij} \cdot {\hat u}_{j})^{2}}
{1 - {\cal X} ({\hat u}_{i} \cdot {\hat u}_{j})} ] \}^{-1/2} . 
\end{eqnarray} 
Here, $\sigma_{0}$ is constant, which corresponds to the width of 
molecules, ${\cal X}$ is a measure of the nonsphericity of the molecule 
defined as ${\cal X} = (k^{2}-1)/(k^{2}+1)$ with $k$ being the 
length-to-width ratio of the molecule.

We use the standard constant-pressure (const-NPT) Monte Carlo simulations to 
obtain the equation of state $P(\rho)$ ( for hard potential, temperature $T$ 
is not present explicitly ). The simulations are performed with $N=108$ or 
$256$ for the molecular elongations $k$ being between $2.0$ and $3.0$. For 
$k \ge 3.0$, a great number of simulations have shown that there is 
isotropic-nematic phase 
transition in the bulk~\cite{Miguel2003,Miguel2001,Padilla1997}, but for 
$k \le 2.0$, the nematic phase is instable in bulk up to a typical high 
density of liquids~\cite{Rigby1989}. 
In simulations of bulk systems, the simulating box is cubic and equally 
fluctuates in three directions, cubic periodic boundary conditions are used. 
In confinement, the distance between two walls ($z$ direction) is fixed, the 
box equally fluctuates only in the two directions ($x$ and $y$) and uses 
periodic boundary conditions. The simulations are organized in MC cycles, 
each MC cycle consisting (on average) of $N$ trial translational and 
rotational molecular displacements and one trial volume fluctuation. The 
maximum step length of each trial move is automatically chosen at each 
pressure for making the acceptable probability fall between $0.4$ and $0.6$. 
The starting configuration is a fcc lattice ( in confinement, 
the number of lattic in $z$ direction may be smaller than that in other 
directions ), which melte at low pressure and equilibrate for 100,000 MC 
cycles. The system is slowly compressed in small pressure steps. For any 
given pressure, the system is typically equilibrated for 100,000 MC cycles 
and an average is taken over 100,000 additional MC cycles, then the final 
configuration is set as the starting configuration of the next pressure. 
But 300,000 MC cycles are used near the isotropic-nematic transition. After 
arriving at a high pressure point, in some cases, we slowly expand the system 
from this high pressure point to a low pressure point.
In calculating average values (such as density, orientational order 
parameter ), we sample partial equilibrated configurations. However, the 
obtained results (average values and their fluctuations) are almost the 
same as that using all equilibrated configurations.

The orientational order parameter $S$ is calculated in the simulations as 
the largest eigenvalue of the ordering $Q_{\alpha \beta}$ tensor, defined 
in terms of the components of the unit vector $u_{i \alpha}$ along the 
principal axis of the molecules,
\begin{eqnarray}
Q_{\alpha \beta} = \frac{1}{N} \sum_{i=1}^{N} ( {3 \over 2} u_{i \alpha} 
u_{i \beta} - {1 \over 2} \delta_{\alpha \beta} )
\end{eqnarray}

In the simulation of confined systems, a similar hard potential is set as 
the interaction between molecule and walls~\cite{Gruhn1998}, thus
\begin{eqnarray}
U^{[k]}(z^{[k]}_{i}, {\hat u}_{i}) = 
\left\{ 
\begin{array}{@{\,}ll}
\infty, & if \ \mbox{$ z^{[k]}_{i} \le \sigma_{w}({\hat u}_{i}) $} \\
0, & if \ \mbox{$ z^{[k]}_{i} > \sigma_{w}({\hat u}_{i}) $},
\end{array}
\right.
\end{eqnarray}
where $z^{[k]}_{i}$ is the distance between center-of-mass of molecule 
$i$ and wall $k$, and $\sigma_{w}$ is given by
\begin{eqnarray}
\sigma_{w}({\hat u}_{i}) = \frac{\sigma_{0}}{2 \sqrt{1-\eta 
|{\hat u}_{i} \cdot {\hat z}|}}.
\end{eqnarray}
Here $\eta = (k^{2}-1)/k^{2}$. We suppose the walls are at $z=0$ and 
$z=D$, respectively. Of course, the center-of-mass of the molecules 
must be between two walls.

In the remainder of this paper, $\sigma_{0}$ is set as unit of length, 
$P/k_{B}T$ is noted as $P$. 

\section{Results}

We first simulate the equation of state of HGO systems with $k=3.0$ in bulk.
For this case, isotropic-nematic (IN) phase transition had been found in 
previous simulations~\cite{Miguel2003,Miguel2001,Padilla1997}. Based on 
free energy calculations of isotropic and nematic phases, the IN transition 
pressure was about $5$ for $k=3.0$. In this paper, we do not try to determine 
exactly IN transition point by using these free energy methods, but only 
estimate the transition from their $S-P$ curves. 

\begin{figure}
\centerline{\epsfxsize=10cm \epsfbox{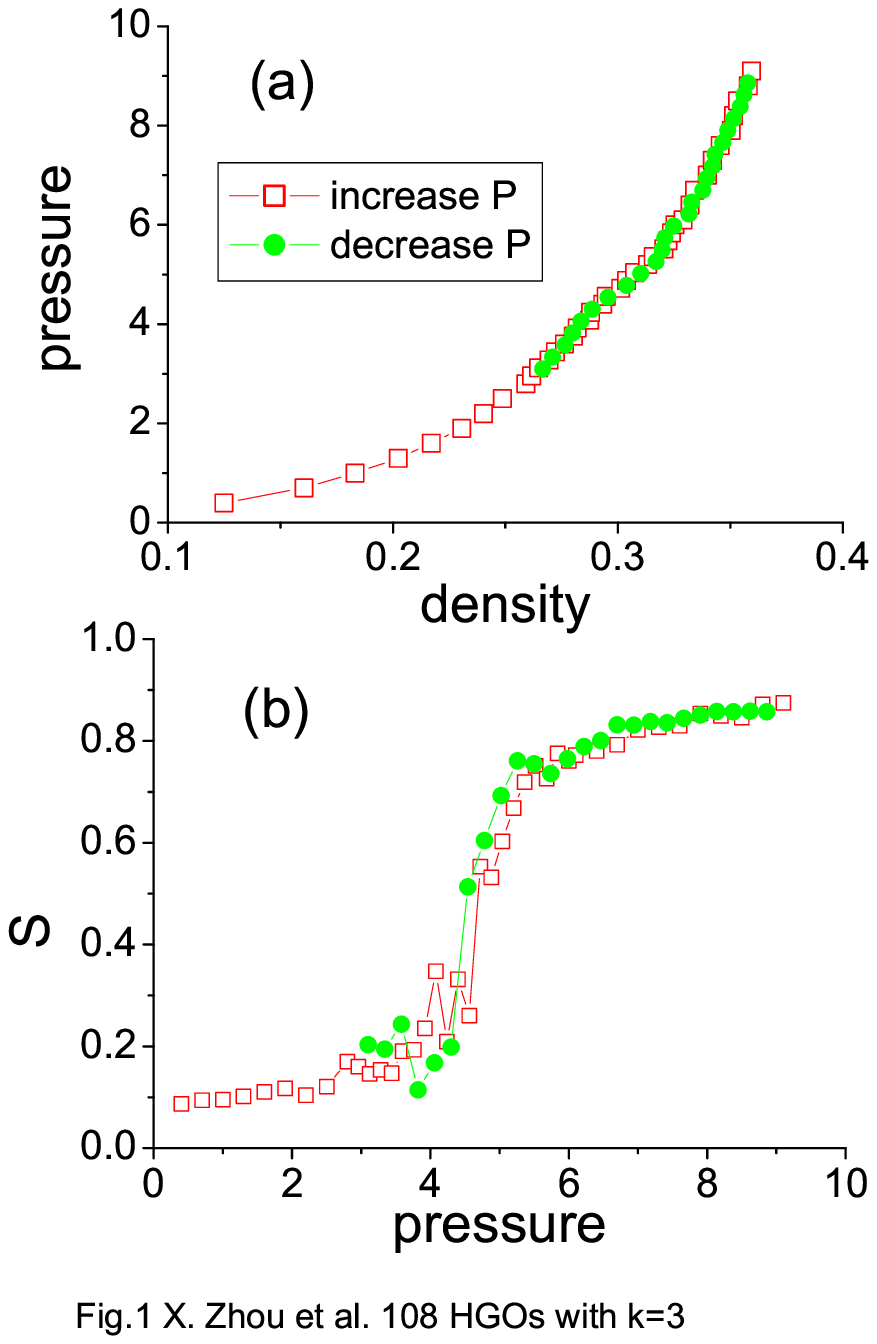}}
\caption{ Isotropic-nematic phase transition in bulk is shown. Here, the 
number of 
molecules ($N$) in our simulations is $108$. The anisotropic parameter $k=3$.
(a) Equation of state of the system $P(\rho)$ is shown. (b) Orientation 
order parameter $S$ versus the employed pressure. 
Shown simulation results include that an increasing pressure and a decreasing 
pressure process. 
\label{fig1}}
\end{figure}

Our results are shown in Fig.(\ref{fig1}), It is clear, the equations of 
state exhibit a discontinuity at $P \sim 5$. In Fig. (\ref{fig1}) (${\bf b}$), 
the behavior of the orientation order parameter $S$ shows that a 
spontaneous IN phase transition occurs at the discontinuous point. The result 
in increasing and decreasing pressure (compressing and expanding simulations) 
are similar to each other and in agreement with previous simulations. 

\begin{figure}
\centerline{\epsfxsize=10cm \epsfbox{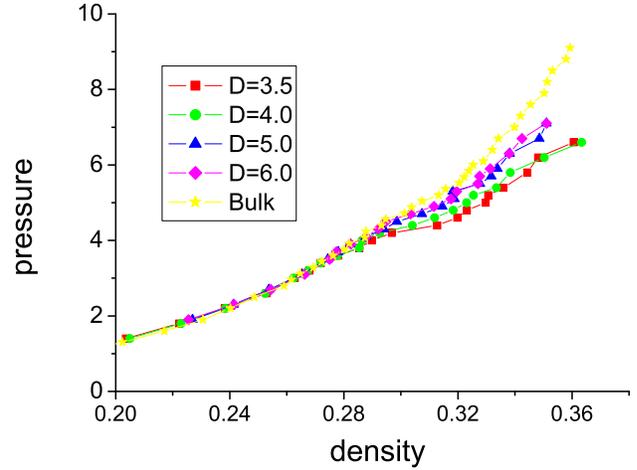}}
\caption{  Equation of state $P(\rho)$ of a HGO system in slit pores is shown. 
Here, $N = 256$, and $k=3$. $D$ is the width of the used slit pore in our
Monte Carlo simulations.
\label{fig2}}
\end{figure}

\begin{figure}
\centerline{\epsfxsize=10cm \epsfbox{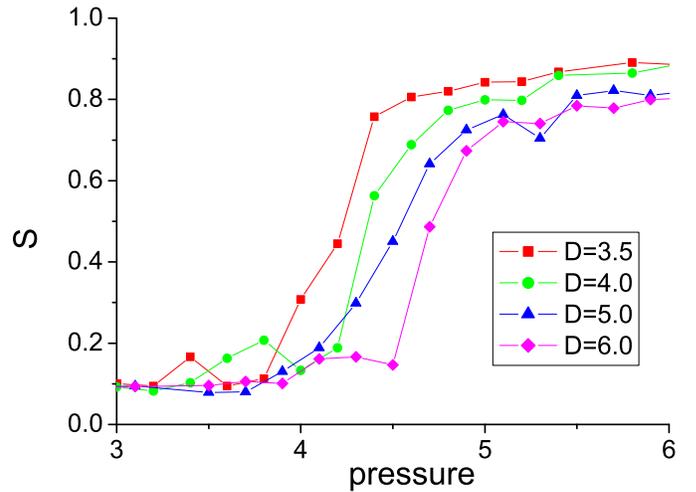}}
\caption{  Orientation order parameter $S$ versus 
pressure $P$ in slit pores with different $D$. A shift in IN 
transition is clearly shown. Here, $N=256$ and $k=3$.
\label{fig3}}
\end{figure}

\begin{figure}
\centerline{\epsfxsize=10cm \epsfbox{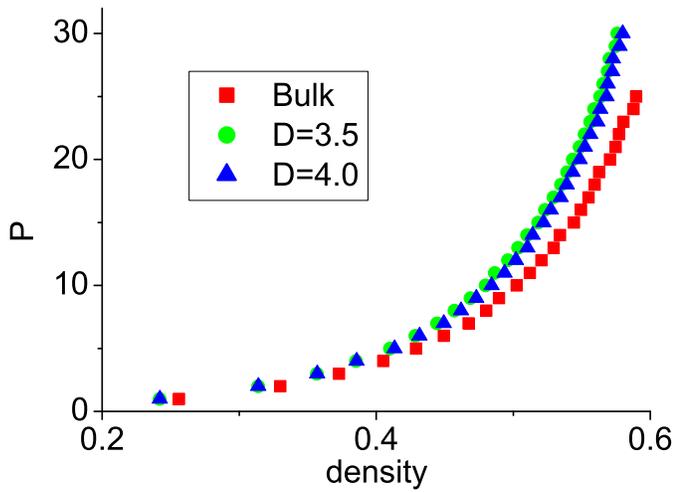}}
\caption{  Equations of state $P(\rho)$ of a HGO system in bulk and in 
confinement are shown. Here $N= 256$ and $k=2$. The density of the system 
is increased to a typical high density of liquids from zero.
\label{fig4}}
\end{figure}

\begin{figure}
\centerline{\epsfxsize=10cm \epsfbox{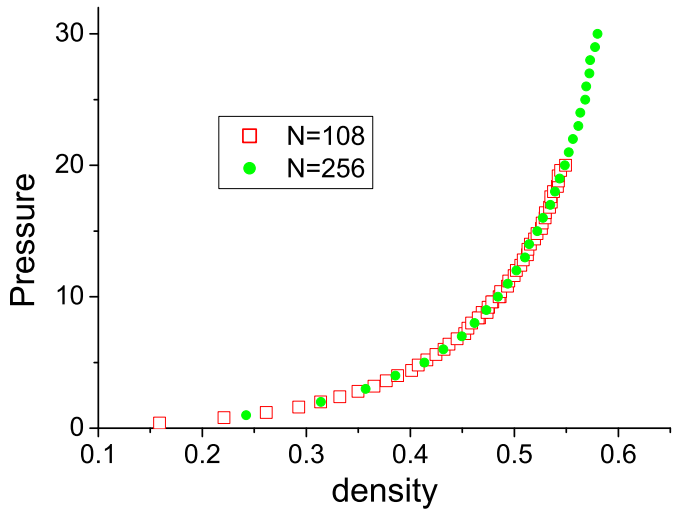}}
\caption{  Equations of state $P(\rho)$ of systems are shown. Here $N = 256$ 
and $108$. $k=2$ and $D=4.0$. Our result indicates no obvious sized effects 
in our simulations of confined systems.
\label{fig5}}
\end{figure}

\begin{figure}
\centerline{\epsfxsize=10cm \epsfbox{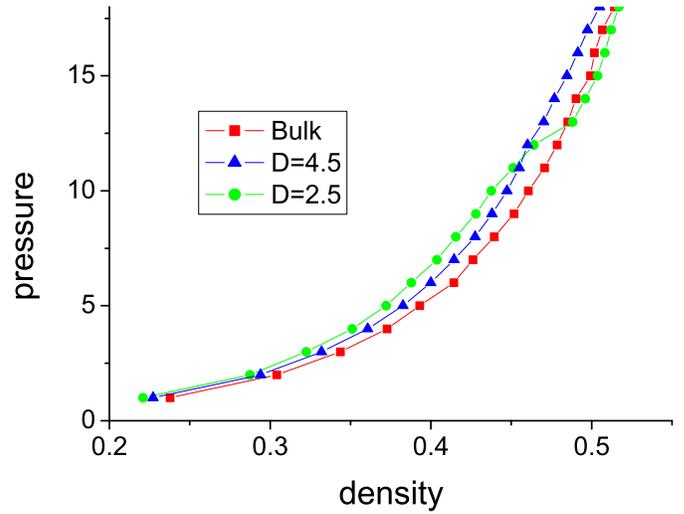}}
\caption{  Equations of state $P(\rho)$ in a system in bulk and in confinement
are shown, respectively. Here, $N = 108$ and $k=2.2$. The densities of the 
system are increased to a typical high density from zero. In a thin slit 
pore ($D=2.5$), the equation of state exhibits a discontinity.
\label{fig6}}
\end{figure}

\begin{figure}
\centerline{\epsfxsize=10cm \epsfbox{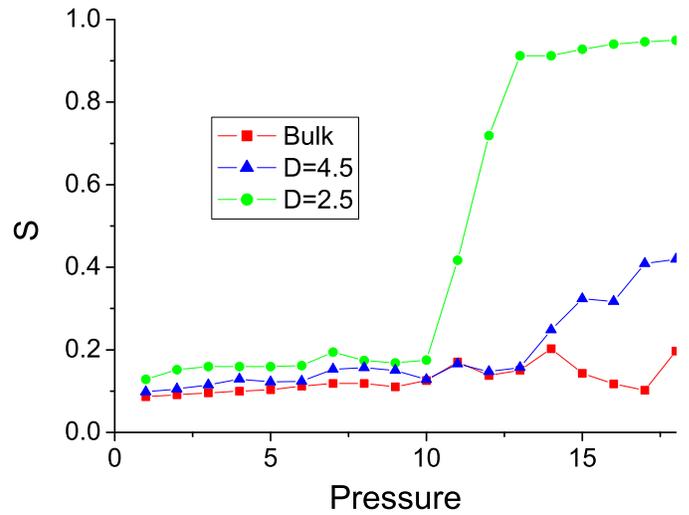}}
\caption{  Orientation order parameters $S$ versus 
pressure $P$ in a HGO system in bulk and in slit pores, where $N = 108$ and 
$k=2.2$. An IN transition is found in $D=2.5$, but the transition does not 
exist in bulk. 
\label{fig7}}
\end{figure}

To test the effects of confinement, we study the confined HGO systems with
$k=3.0$ in different $D$, the obtained results are shown in Fig.(\ref{fig2}) 
and Fig.(\ref{fig3}). This leads to our first interesting result: the IN 
transition point shift to lower pressure (i. e. density) direction as the 
distance between two walls $D$ decreases. Usually, for small anisotropic 
molecules, the nematic phase is instable, the reason can be understood. The 
system will freeze or crystallize before forming liquid-crystal phase while 
increasing the pressure or decreasing the temperature. So the shift at low 
pressure indicates that nematic phase form in confinement easiler than in 
bulk. Therefore, we suspect the required smallest $k$ forming nematic 
phase is smaller in confinement than that in bulk. 
In Fig.(\ref{fig4}) and Fig.(\ref{fig5}), we show the equation of state of 
smaller anisotropic molecules ($k=2.0$). 
Up to a typical high density of liquid ( packing fraction $y$ is about $0.5$), 
nematic phase is not found in bulk. Here $y= \rho v_{m}$, where $\rho$ is 
density of the liquid, $v_{m} = \pi/6 k \sigma_{0}^{3}$ is the volume of the 
molecule. The result is in agreement with previous 
simulations~\cite{Rigby1989}. However, we do not find stable nematic phase 
in the system in confinement (the distance $D$ between walls is $3.5$ and 
$4.0$). Consider that $k=2.0$ may be too small, we simulate another system 
with a little larger elongation ($k=2.2$) and confined in a thiner pore 
($D=2.5$), the results are shown in Fig.(\ref{fig6}) and Fig.(\ref{fig7}). 
The system is compressed up to a typical high density ( $y = 0.57$ ), no 
phase transition is found in bulk. But for the confined system in slit pore 
with a part of $D=2.5$, an obvious IN phase transition is found. In 
principle, to strictly determine the existence of liquid-crystal 
phase in bulk and in confinement, we should use free energy methods to 
calculate the global phase diagram of the 
system including calculations of chemical potetials in isotropic phase, 
orientation order phase and crystallized phase, respectively. 
However, in this paper, we use a rougher method used by earlier 
Rigby~\cite{Rigby1989}: compress the system to a typical high density ($y$ 
is about $0.5$), if there is no nematic phase in the density zone, the 
nematic phase does not exist, since in higher density zone, fluids usually 
freeze or crystalize. The result is a rough approximation, but we still can 
obtain a general qualitative conclusion: confinement will decrease the 
required molecular elongation $k$ to form a nematic phase. We believe that a 
nematic phase exists for a certain $k$ in systems in suitable confinement, 
though the nematic phase of the systems is instable in bulk.
Of course, it is important to obtain quantitative results on this effect, 
and we shall study global phase diagrams of more realistic small-anisotropic 
molecules in confinement.

\section{Conclusion}

From the ongoing investigation, we can state that orientation ordered phase
exists in bulk for some values of the molecular elongation $k$ above
a critical value $k_{c}$; for smaller value of $k$, the ordered phase becomes
instable, but as these small $k$ molecules are confined in very thin pores, 
the orientation ordered phase can be found.
This result confirms our expectation that small-anisotropic molecules form 
liquid crystals in confinement. 


Usually, the elongation of liquid 
crystal molecules is about $4-5$ and $k$ of nonspheric inorganic molecules, 
such as $CO_{2}$, $CS_{2}$ etc., are only about $2$. Hence, the usual 
small molecules cannot form liquid-crystal phases in bulk. But in confinement, 
according to our results, the required $k$ forming liquid crystals will 
decrease, so it is interesting to study whether or not some small molecules 
will form liquid crystal phase in special confinement. Due to the wide 
interests in small molecule liquid crystals, it may be important to study 
phase behavior of real confined small molecules. 
On the other hand, complicated geometrical pores may induce more complex 
behavior in fluids~\cite{Gelb1999}. Recently, Chiccoli {\it et al.} studied 
the properties of liquid crystals with dispersed polymer 
fibrils~\cite{Chiccoli2002}, where the polymer fibrils formed networks and 
served as complicate pores to confine liquid crystals. Considering the 
required sizes of pores to induce possible liquid crystal phases in small 
molecules may be very small (in molecule scale), we shall study the 
effects of confinement in small anisotropic molecules with dispersed large 
anisotropic molecules. 
In the mixture, large anisotropic molecules may serve as complex pores to
confine small molecules and induce possible orientation-ordered phases, 
which might induce characteristic of small-molecule liquid crystals. 

X. Z is financially supported by the Grants-in-Aid for 
Scientific Research of JSPS; H. Chen is supported by the Singapore 
Millennium Scholarship. X. Z extend gratitude to Dr. T. Manaka and Mr. 
C.-Q. Li.

(Correspondent address: zhou@pe.titech.ac.jp).

\end{document}